\documentclass{PoS}
\usepackage{graphicx}
\usepackage{wrapfig}
\usepackage{subcaption}
\usepackage{xfrac}
\usepackage{comment}
\usepackage{tabularx}
\usepackage{overpic}
\usepackage{url}
\usepackage[numbers]{natbib}
\setcitestyle{open={[},close={]},citesep={,}} 

\usepackage[utf8]{inputenc}

\title{CTLearn: Deep Learning for Gamma-ray Astronomy}

\ShortTitle{CTLearn: Deep Learning for Gamma-ray Astronomy}

\author{\speaker{D. Nieto}$^{1}$, A. Brill$^{2}$,  Q. Feng$^{2}$, T. B. Humensky$^{2}$, B. Kim$^3$, T. Miener$^{1}$, R. Mukherjee$^{4}$, J. Sevilla$^{5}$
{\footnotesize
\\$^{1}$Instituto de Física de Partículas y del Cosmos and Departamento de EMFTEL, Universidad Complutense de Madrid, Madrid, Spain,
$^{2}$Columbia University, Department of Physics, New York, NY, USA,
$^{3}$University of California Los Angeles, Division of Astronomy and Astrophysics, Los Angeles, CA, USA,
$^{4}$Department of Physics and Astronomy, Barnard College, Columbia University, NY 10027, USA\\
$^{5}$Facultad de Ingeniería Informática, Universidad Complutense de Madrid, Madrid, Spain,

E-mail: \email{d.nieto@ucm.es}}}

\abstract{CTLearn is a new Python package under development that uses the deep learning technique to analyze data from imaging atmospheric Cherenkov telescope (IACT) arrays.  IACTs use the Cherenkov light emitted from air showers, initiated by very-high-energy gamma rays, to form an image of the longitudinal development of the air shower on the camera plane. The spatial, temporal, and calorimetric information of the originating high-energy particle is then recorded electronically. The sensitivity of IACTs to astrophysical sources depends strongly on the efficient rejection of the background of much more numerous cosmic-ray showers. CTLearn includes modules for running machine learning models with TensorFlow, using pixel-wise camera data as input. Its high-level interface provides a configuration-file-based workflow to drive reproducible training and prediction. We illustrate the capabilities of CTLearn by presenting some results using IACT simulated data. }

\FullConference{36th International Cosmic Ray Conference -ICRC2019-\\
		July 24th - August 1st, 2019\\
		Madison, WI, U.S.A.}
		
\begin{document}

\section{Introduction}

In this contribution we introduce CTLearn~\cite{ctlearn}, a new Python package under development that enables the application of deep learning techniques to event reconstruction in the analysis of data from imaging atmospheric Cherenkov telescope (IACT) arrays. IACTs are instruments sensitive to very-high-energy ($\gtrsim 10$'s GeV) gamma rays, by capturing images of the extended air showers that these gamma rays, as well as cosmic rays, produce when absorbed by the atmosphere. The information contained in those images can be exploited to infer the properties of the shower progenitor: particle type, energy and arrival direction. This so-called event reconstruction is crucial in the analysis of IACT data, since it determines the effectiveness of the background suppression (the rejection of the much more frequent cosmic-ray initiated showers) and the angular and energy resolutions of the instrument, the main drivers of IACT sensitivity to astrophysical gamma-ray sources.

The morphological differences between gamma-ray and cosmic-ray initiated showers, translated into their IACT images, can be used to distinguish them. Handcrafted features extracted from the images and box cuts over the multidimensional space of features were originally used for particle classification~\cite{1985ICRC....3..445H}, later evolving into more sophisticated strategies where supervised learning algorithms like Random Forests~\cite{2008NIMPA.588..424A} or Boosted Decision Trees~\cite{2009APh....31..383O,2011APh....34..858B,krause2017improved},  trained on those handcrafted features, substantially improved the classification performance and, consequently, the sensitivity of the IACTs that implemented them into their analysis chains. Deep convolutional neural networks (DCNs), a particular class of deep learning algorithms, currently are the most successful machine learning method for computer vision, excelling at image classification among other tasks~\cite{Goodfellow-et-al-2016}. DCNs belong to the class of representation learning, where the agent that crafts the features to learn from is the algorithm itself. DCNs present the potential to access all the information contained in the images, not only that one condensed in handcrafted features extracted from those images. The capability of DCNs to classify gamma-ray from cosmic-ray simulated events was demonstrated in~\cite{2017arXiv170905889N} for the first time. Their ability to tag muon events was later shown in~\cite{veritas_cnn}, as well as their potential to reconstruct the energy and arrival direction of simulated gamma-ray events~\cite{2018arXiv181000592M} and to improve the sensitivity of the analysis on real data~\cite{2019APh...105...44S}.

CTLearn aims to help the IACT community explore and use deep learning models, with a particular focus on DCN-based models. This contribution is structured as follows: Sec.~\ref{sec:design} presents the framework design for this package, discusses the data input method, the configurable settings, and describes the currently implemented models for gamma-ray/cosmic-ray particle classification; Sec.~\ref{sec:benchmark} contains benchmarking results for the models introduced in the previous section, obtained after training on a dataset of Monte Carlo simulated IACT events; and we conclude by briefly describing plans for future developments in Sec.~\ref{sec:conclusion}.

\section{CTLearn framework}
\label{sec:design}

CTLearn provides a backend for training deep learning models for IACT event reconstruction using TensorFlow~\cite{tensorflow}. CTLearn allows its user to focus on developing and applying new models while making use of functionality specifically designed for IACT event reconstruction. It uses YAML configuration files to encourage reproducible training and prediction, ensuring that settings used to train a model are explicitly set and automatically recorded. Data loading and pre-processing are performed using an associated external package, DL1-Data-Handler (DL1DH,~\cite{dl1dh}). DL1DH contains modules for writing data to an HDF5 format designed for IACTs using PyTables, loading data stored in this format, applying arbitrary transformations to the data, and mapping hexagonally spaced pixels to rectangular matrices.

\begin{figure}
    \centering
    \includegraphics[width=0.9\textwidth]{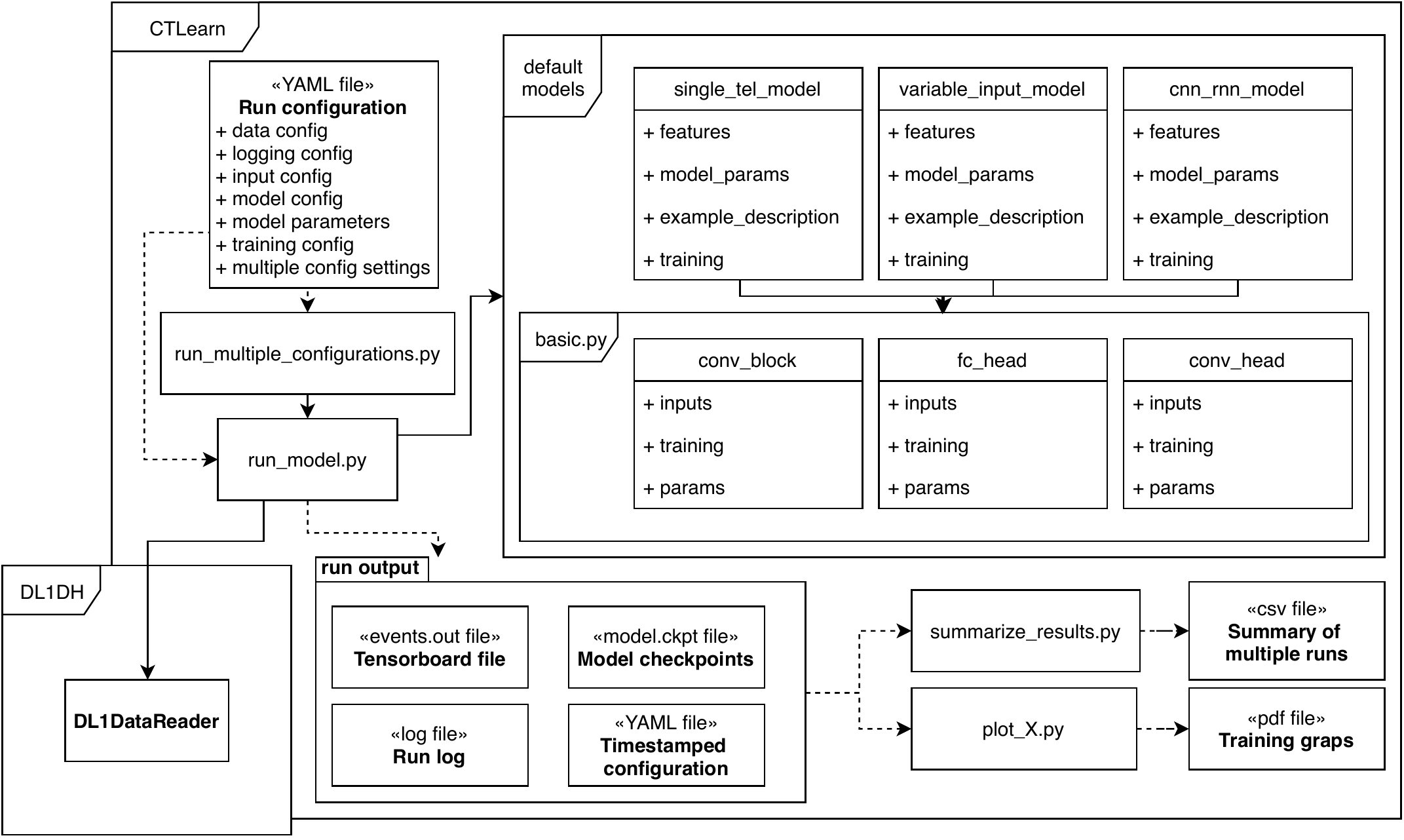}
    \caption{Diagram summarizing CTLearn's framework design.}
    \label{fig:uml}
\end{figure}

CTLearn is organized around the \texttt{run\_model} module, which parses the configuration, loads the data, and initializes the model; then, depending on the mode, it either trains the model, uses the trained model to generate predictions on a test set, or displays properties of the dataset. CTLearn also includes a number of ancillary scripts, providing a convenient way to summarize the results and make plots using the output it produces. A diagram summarizing the CTLearn framework can be found in Fig.~\ref{fig:uml}.

The version of the package at the time of these proceedings is v0.4.0, running on Python 3.7.3 and TensorFlow 1.13. Other dependencies are DL1-Data-Handler, NumPy, and PyYAML, in addition to Matplotlib, Pandas, and Scikit-Learn, which are used only in the supplementary scripts.

\subsection{Settings}

As mentioned above, CTLearn keeps all run settings in a single configuration file. A description of the main configuration options follows.

{\bf Data input} The user can describe the dataset to use and relevant settings for loading and processing it. These settings are used to initialize a DL1DH DL1DataReader, which loads the data files, maps the images from vectors to arrays, applies preprocessing, and returns the data as an iterator. DL1DH includes a number of different image mapping methods \cite{2019-icrc-hex}. Data can be loaded in three modes: mono, where single images of one telescope type are loaded; stereo, where events (that is, the collection of images from all telescopes triggered by a given air shower) of one telescope type are loaded; and multi-stereo, where events including multiple telescope types are loaded.

{\bf Model}
CTLearn works with any TensorFlow model obeying a generic signature \texttt{logits = model(features, params, example\_description, training)} where \texttt{logits} is a vector of raw (non-normalized, pre-Softmax) predictions, \texttt{features} is a dictionary of tensors, \texttt{params} is a dictionary of model parameters, \texttt{example\_description} is a DL1DataReader example description, and \texttt{training} is a Boolean that's true when training and false when validating or testing.
In addition, CTLearn includes three models for gamma-ray/cosmic-ray classification. The {\em single-tel} model classifies single telescope images using a convolutional network, whereas the {\em CNN-RNN} and {\em variable input} models perform array-level classification by feeding the output of a DCN for each telescope into either a recurrent network, or a convolutional or fully-connected network head, respectively. All three models are built on a simple, configurable module called \texttt{Basic}, providing customizable models essentially consisting of stacks of convolutional layers. These models use rectified linear units as the activation functions in all layers.

{\bf Training} The user can customize training hyperparameters such as the fraction of data randomly extracted from the training dataset for validation purposes, the number of validations to run, how often to evaluate on the validation set, the optimizer, and the base learning rate for the chosen optimizer. Loss class weighting, for unbalanced datasets, is also available. 

{\bf TensorFlow} The user can set parameters for data input using the TensorFlow \texttt{Dataset} and \texttt{Estimator} APIs. The TensorFlow debugger can also be optionally invoked. 

{\bf Prediction} The user can specify the prediction settings such as the path to write the prediction file and whether to save the labels and example identifiers along with the predictions.

{\bf Multiple configurations} CTLearn features a tool to run multiple configurations in series, sourcing from a single configuration file. This tool can be used to optimize hyperparameters by running over discrete sets or performing grid or random searches over linear or logarithmic-spaced ranges. 

{\bf Logging} The user specifies the directory to store TensorFlow checkpoints and summaries, a timestamped copy of the run configuration, and optionally a timestamped file with logging output.

\subsection{Some built-in models}
\label{sec:models}
In the following, the {\em single-tel} and the {\em CNN-RNN} models are described in more detail. 

{\bf {\em Single-tel} model}
The {\em single-tel} model consists of four convolutional layers with 32, 32, 64, and 128 filters and a kernel size of 3 in each layer, interspaced by an activation layer followed by a max-pooling layer with a kernel size (and stride) of 2; the output of the convolution block is then flattened and fed to a fully connected layer with an output dimensionality of 2, the number of classes.

\begin{wrapfigure}{r}{0.45\textwidth}
    \centering
    \vspace{-40pt}
    \includegraphics[width=0.45\textwidth]{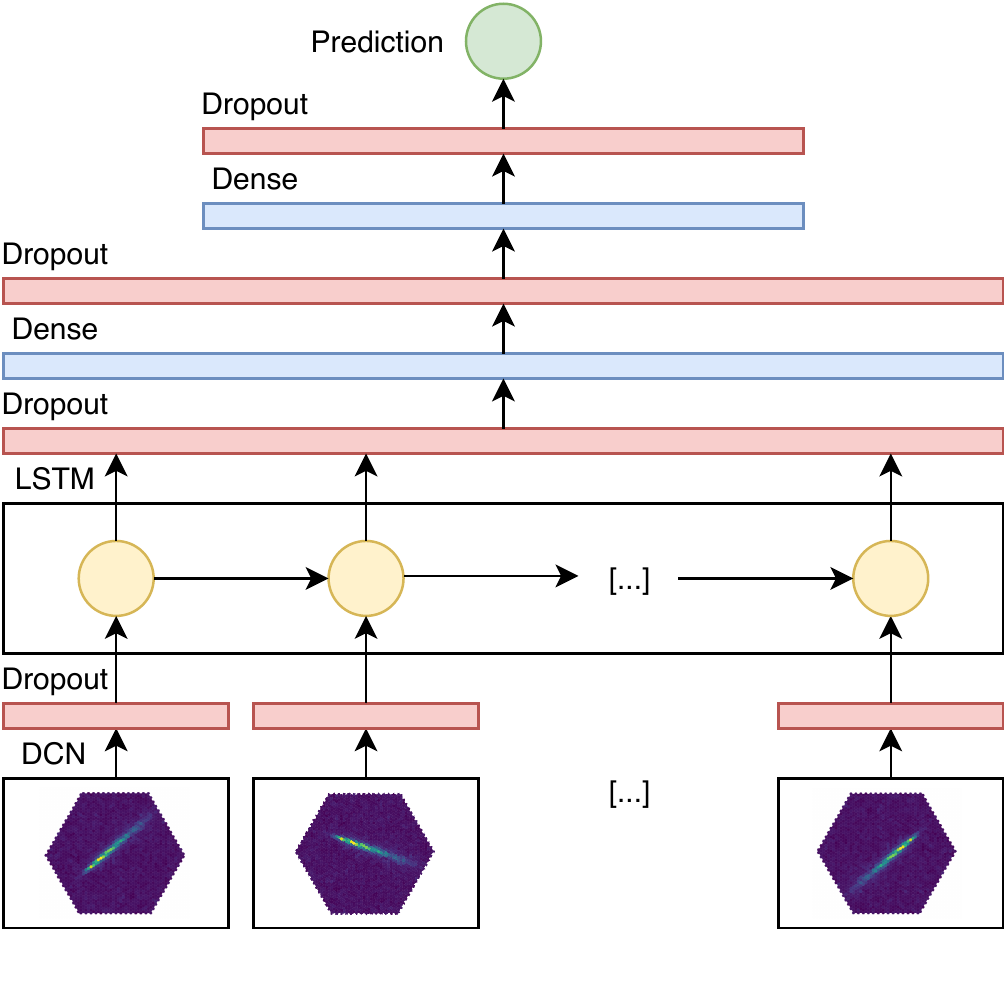}
    \caption{Diagram depicting the main layers of the {\em CNN-RNN} model.}
    \label{fig:cnn-rnn-model}
\end{wrapfigure}

{\bf {\em CNN-RNN} model}
In order to take advantage of the stereoscopic view of the air shower as imaged by an array of telescopes, DCN models must be able to take several images as their input. The {\em CNN-RNN} model (from the combination of convolutional neural network, CNN, and recurrent neural network, RNN), similar to the architecture presented in~\cite{2019APh...105...44S}, takes as input a fixed number of images coming from the same imaged shower; each image is fed into a convolutional block, defined identically to the {\em single-tel} model (weights are shared among all convolutional blocks participating in this stage); the output tensor from each convolutional block is then fed into a dropout layer that in turn feeds into a recurrent neural network, specifically, an LSTM  (long short-term memory) layer, with a size for the hidden state of 2048; after flattening, the output of the LSTM layer is then passed to a concatenation of three fully connected layers, each preceded by a dropout layer, with 1024, 512, and two neurons respectively. The default dropout value for this model is set to 0.5. A diagram depicting the main architecture of the {\em CNN-RNN} model can be found in Fig.~\ref{fig:cnn-rnn-model}.

\section{Benchmark}
\label{sec:benchmark}

\subsection{Benchmark dataset}

The Cherenkov Telescope Array (CTA)\footnote{\href{www.cta-observatory.org}{www.cta-observatory.org}} is the next generation ground-based observatory for gamma-ray astronomy at very-high energies, aiming to improve on the sensitivity of current-generation experiments by an order of magnitude and provide energy coverage from 20 GeV to more than 300 TeV. CTA will consist of two installations, located in the Northern (La Palma, Spain) and Southern (near Cerro Paranal, Chile) Hemispheres, accounting for more than 100 telescopes and allowing for full sky access. The dataset used in this work comes from a reduction of the third large-scale Monte Carlo production for CTA, whose main purpose was to issue a final recommendation for the layout of telescopes that will define both the Northern and the Southern Hemisphere arrays of the observatory~\cite{Acharyya:2019nwy}. This reduction, from raw to calibrated data, was performed on the EGI\footnote{\href{www.egi.eu}{www.egi.eu}} using DL1DH. We restricted ourselves to simulated data from the Southern array, containing four large-size telescopes (LSTs), 25 medium-size telescopes (MSTs), and 70 small-size telescopes (SSTs) arranged in the baseline recommended layout "S8" (following the notation in~\cite{Acharyya:2019nwy}). Out of all the simulated pointing positions, we selected runs with a Zenith angle of 20$^{\circ}$ and an Azimuth angle of 0$^{\circ}$ (North pointing). Concerning the particle type, we considered diffuse, gamma-ray and proton-initiated events in a balanced way, so both populations contribute equally to the statistics of the datasets. The selected dataset was randomly drawn from the source dataset following the described criteria and accounts for almost 400,000 events summing approximately 4 million images. The dataset was split into a training dataset and a test dataset with an 8/2 ratio. We trained the {\em single-tel} and {\em CNN-RNN} models on the following seven telescope designs proposed for CTA: the only model for LST, the two MSTs with Davies-Cotton optics design, equipped with FlashCam or NectarCam cameras (MST-F and MST-N respectively), the dual mirror Schwarzschild-Couder MST (MST-SC), the single-mirror SST equipped with DigiCam camera (SST-1M), and the two dual-mirror SST designs, SST-ASTRI (SST-A) and SST-CHEC (SST-C). More details on the different telescope designs for CTA can be found in~\cite{Acharyya:2019nwy} and references therein.  Table~\ref{tab:stats} gives a more detailed description of the statistics of the dataset.

\begin{table}[t]    
\centering
\resizebox{\textwidth}{!}{
\begin{tabular}{ |c|c|c|c|c|c|c|c|c| }  
\hline
 Events/Images & LST & MST-F & MST-N & MST-SC & SST-1M & SST-C & SST-A & All\\
 \hline
 Training  &  89/187 & 259/770 & 279/891 & 231/626 & 206/440 & 198/440 & 192/472 & 392/3827\\
  Test &  18/39 & 54/160 & 58/185& 48/130 & 43/92 & 41/92&40/98 & 82/796 \\
  \hline
\end{tabular}}
\caption{Statistics of the datasets used in this work. The numbers represent thousands of triggered events and total number of images generated in those events, broken down by telescope type.}
\label{tab:stats}
\end{table}

\subsection{Benchmark results}

\begin{figure}[t]
    \centering
    \includegraphics[width=0.32\textwidth]{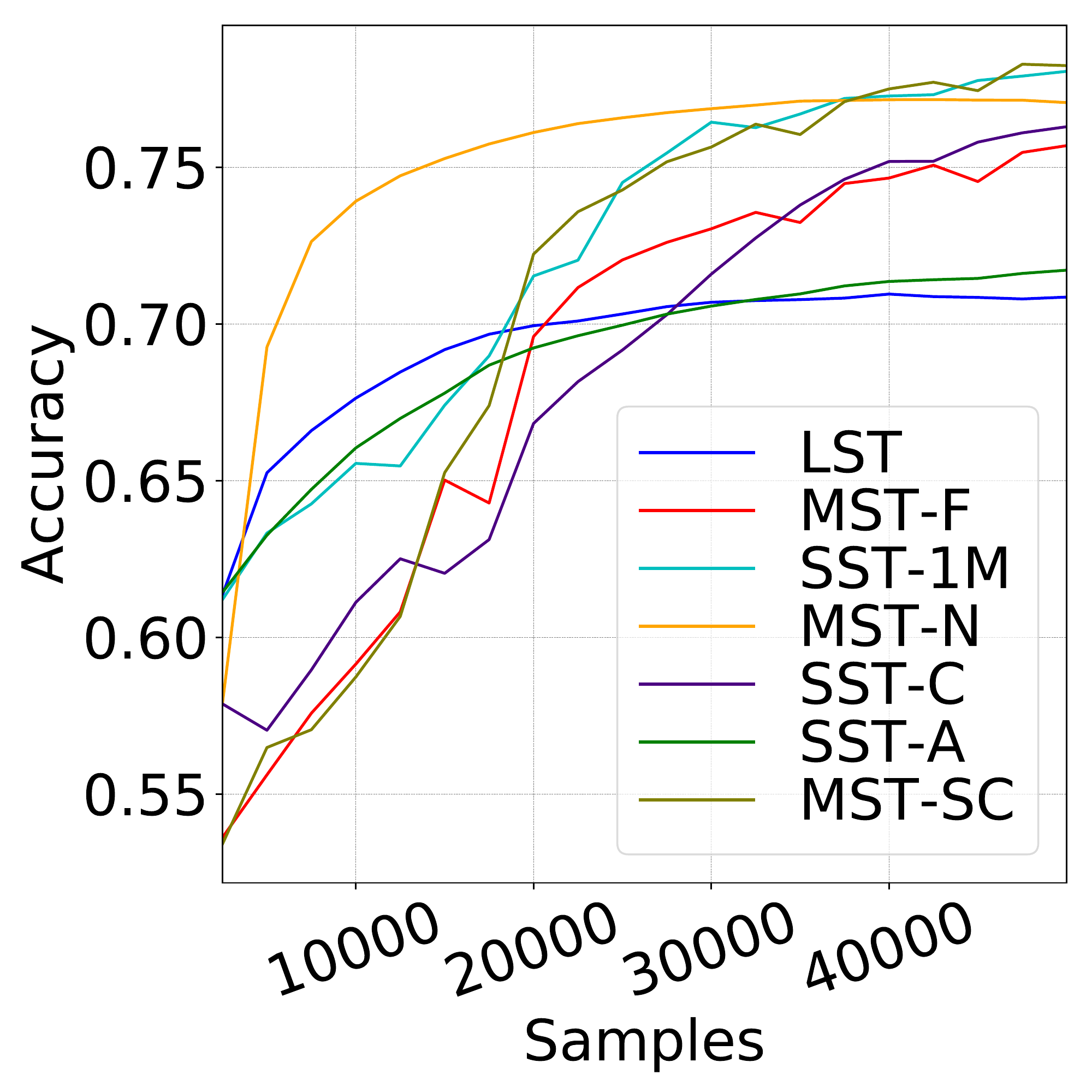}
    \includegraphics[width=0.32\textwidth]{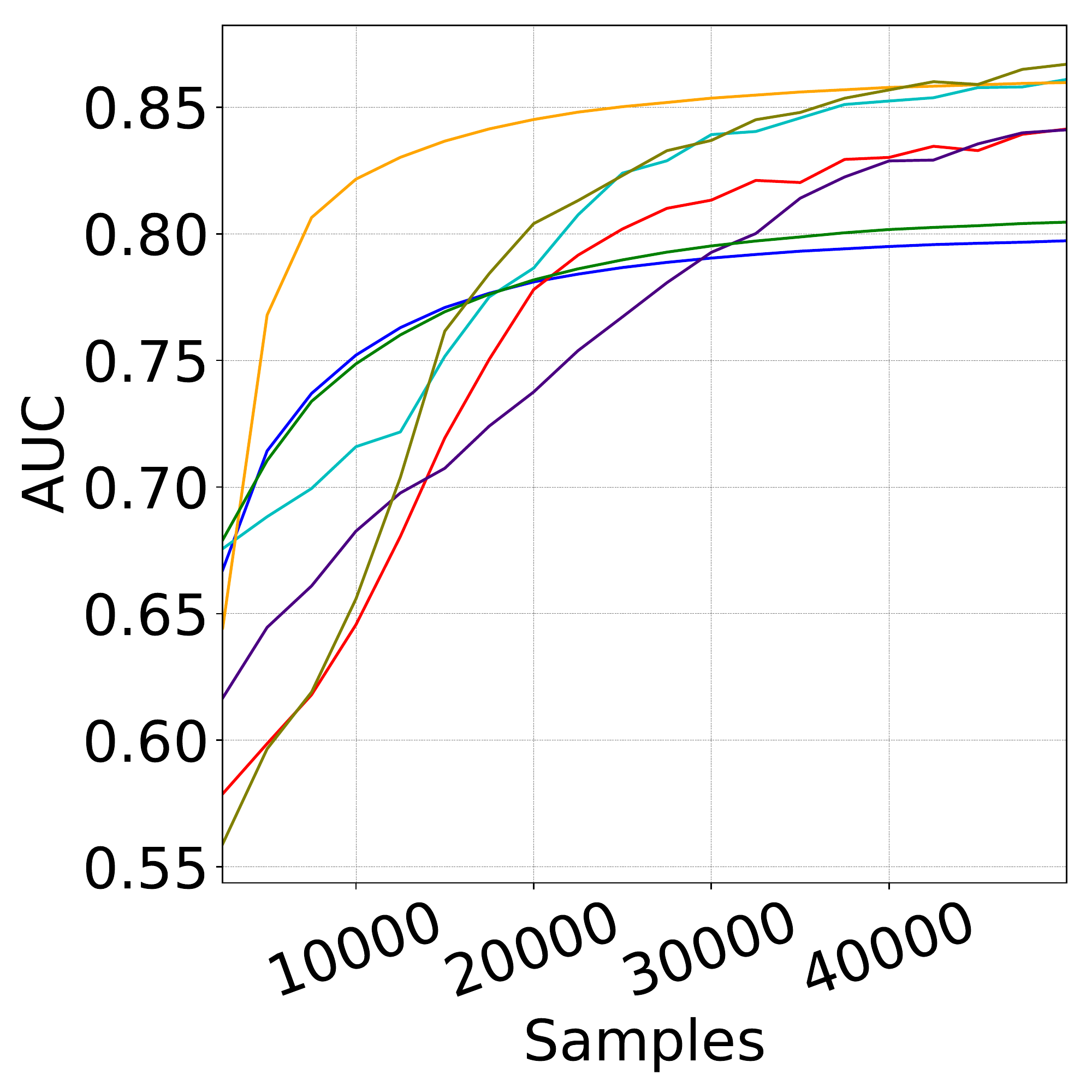}
    \includegraphics[width=0.32\textwidth]{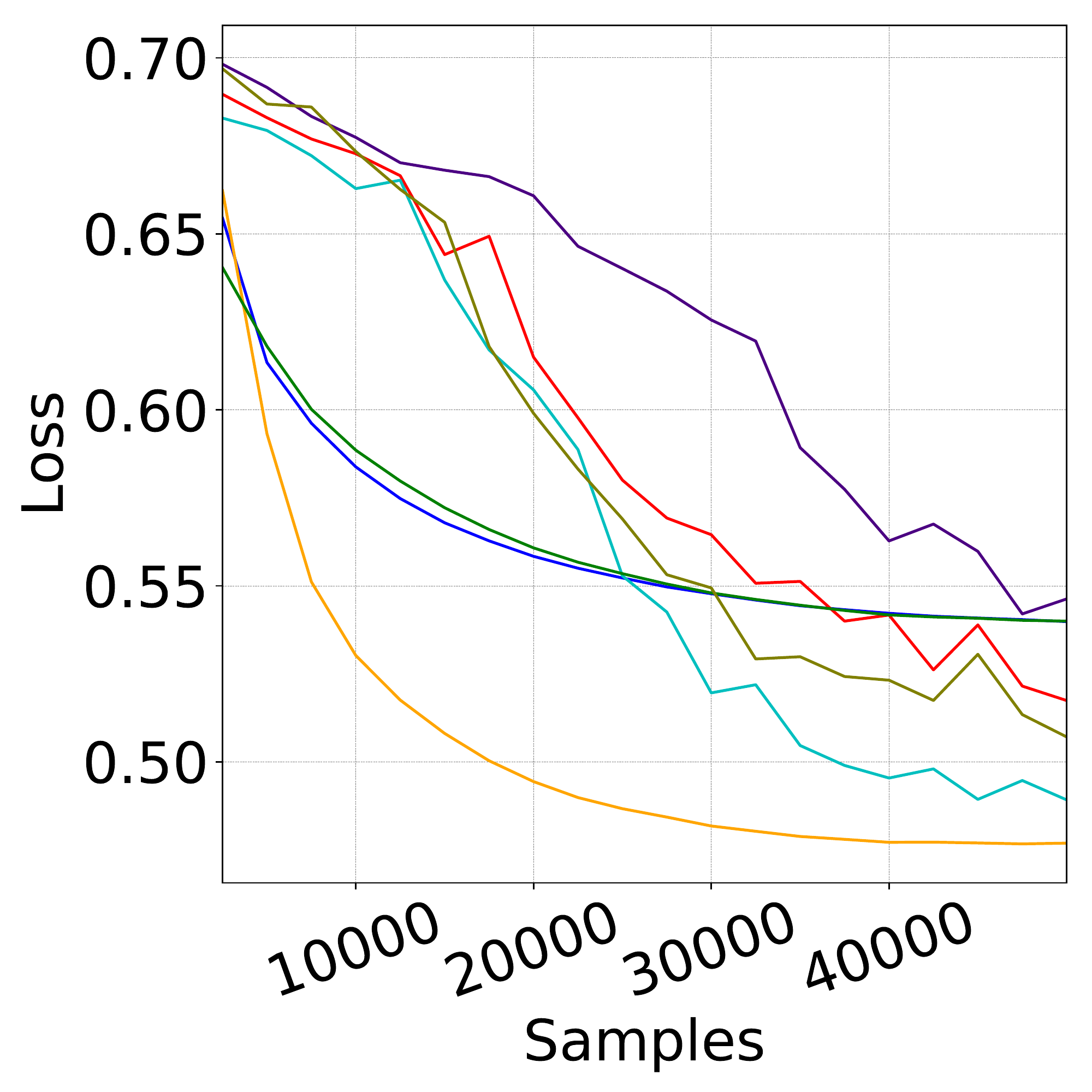}
    \includegraphics[width=0.32\textwidth]{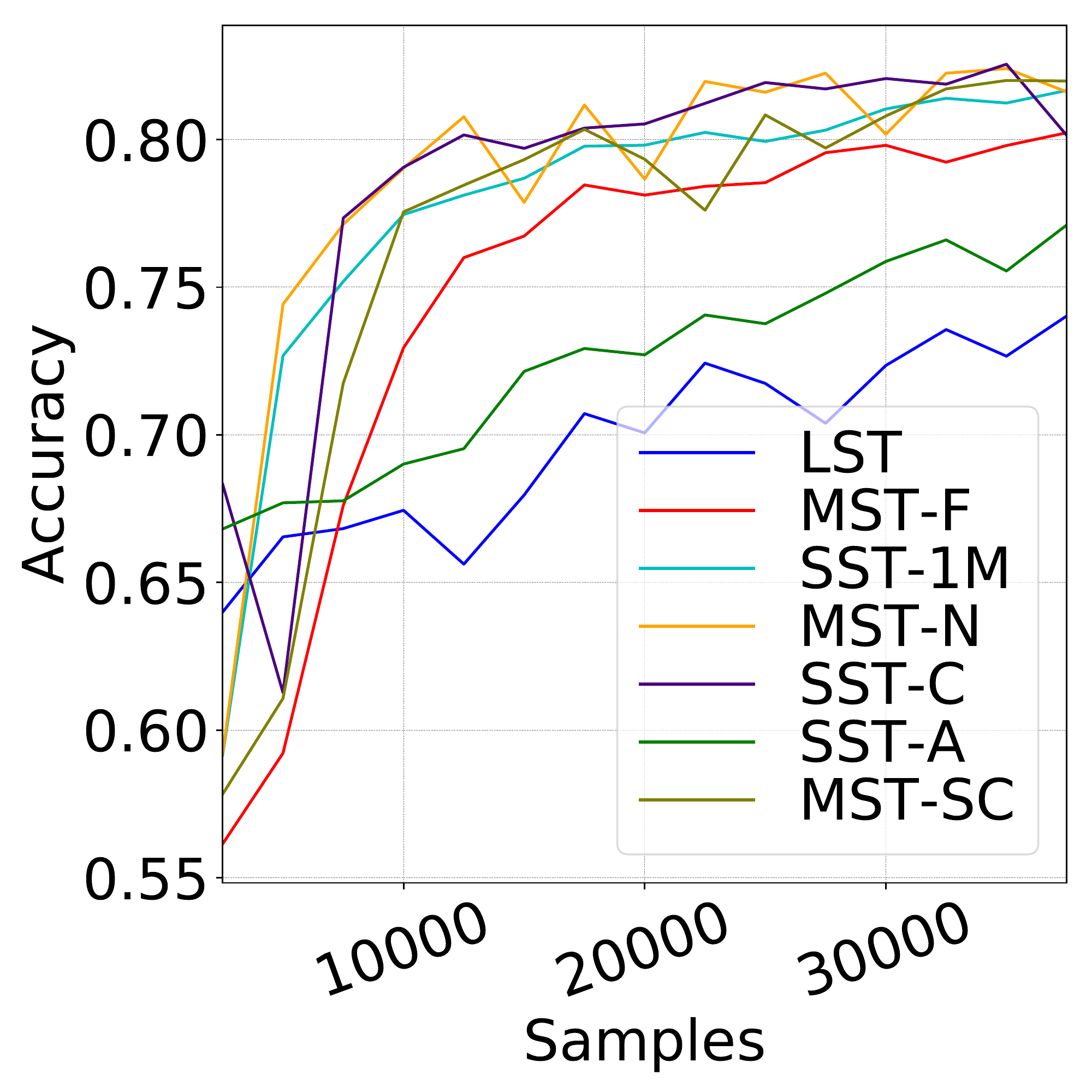}
    \includegraphics[width=0.32\textwidth]{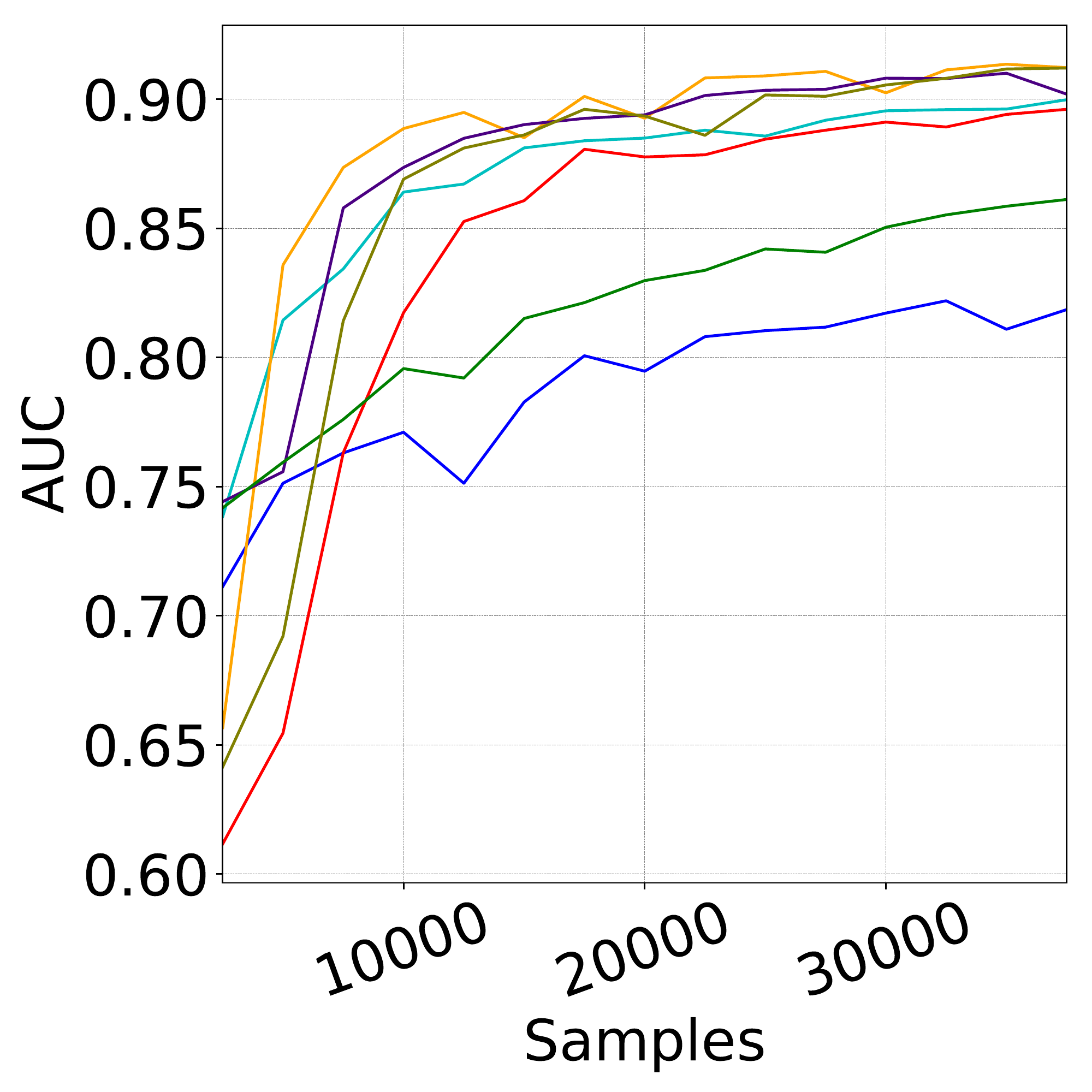}
    \includegraphics[width=0.32\textwidth]{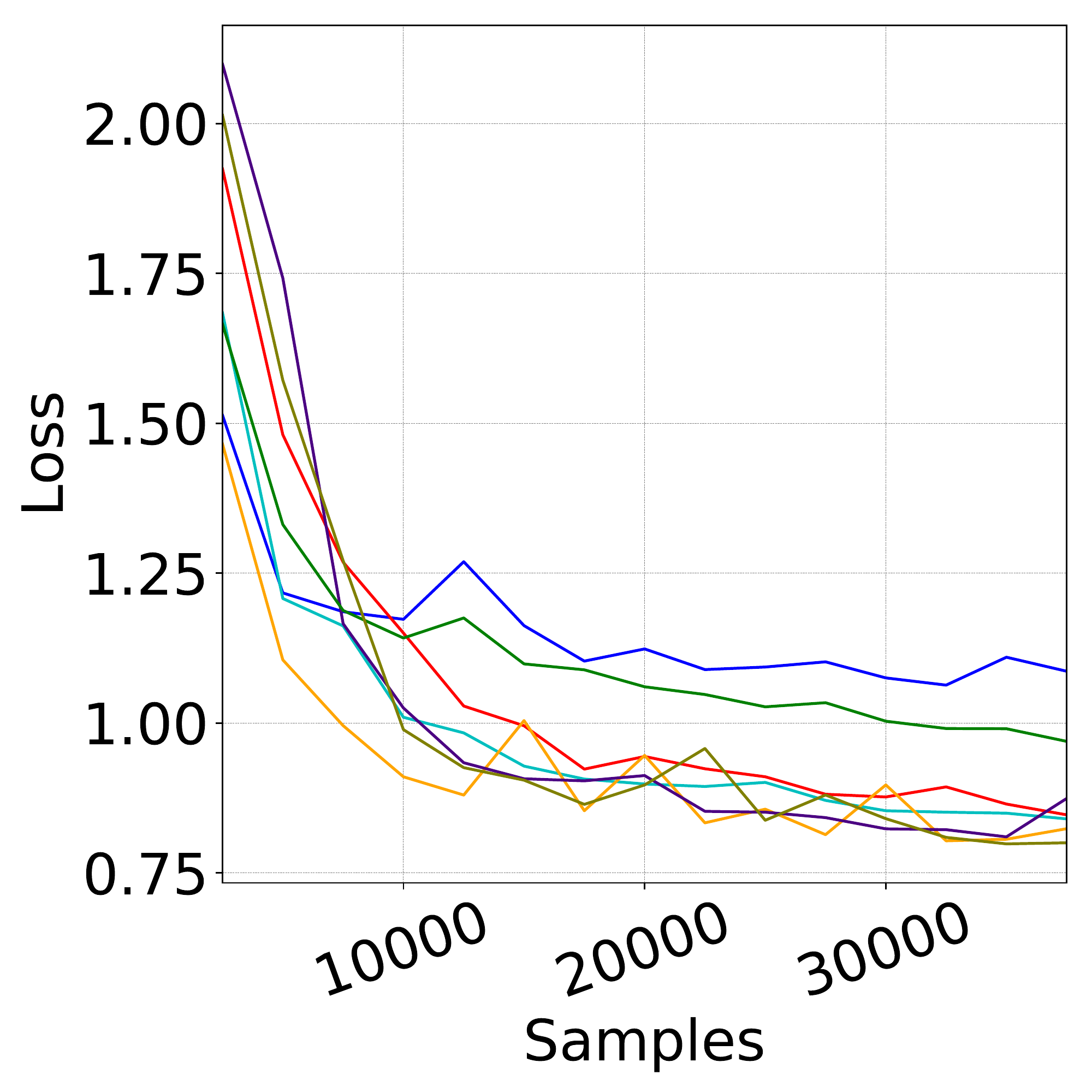}
    \caption{Evolution of the main learning metrics for the {\em single-tel} (top panels) and {\em CNN-RNN} (bottom panels) models as a function of number of samples and telescope type. }
    \label{fig:benchmarks}
\end{figure}

We trained the {\em single-tel} model on 50,000 batches of 64 images each and the {\em CNN-RNN} model on 40,000 batches of 16 events each. Both were validated every 2,500 batches. These settings were chosen to end training approximately when validation loss stops decreasing, and thus provide an illustration of the learning capacities of the models. Images from cameras featuring pixels arranged in hexagonal lattices were mapped to 2D arrays using bilinear interpolation~\cite{dl1dh}. The evolution of the accuracy, AUC, and loss as a function of number of samples run through the model can be found in Fig.~\ref{fig:benchmarks} for both models and telescope type. The accuracy and AUC values for the validation and test sets are summarized in Table~\ref{tab:results_summary}. We found an excellent match between the metrics obtained from the validation and test sets, with the smallest and largest discrepancies being 0.6\% and 1.2\% in AUC for the {\em simgle-tel} model, and 0.1\% and 1.0\% in AUC for the {\em CNN-RNN} model. Test AUC values for the {\em single-tel} model range from 0.78 and 0.81 in the case of the LST and SST-A, respectively, to the 0.84 -- 0.87 range where the rest of the telescope designs are located. Test AUC values for the {\em CNN-RNN} model are located around 0.90 for most telescope designs. 

\begin{table}[t]    
\centering
\resizebox{0.9\textwidth}{!}{
    \begin{tabular}{|c|c|c|c|c|c|c|c|c|} 
        \hline
        \multicolumn{2}{|c|}{{\em Single-tel} model} & LST& MST-F & MST-N & MST-SC & SST-1M & SST-C & SST-A \\
        \hline
        Validation& Acc& 0.701&0.762&0.784&0.795&0.781&0.753&0.733\\
        & AUC &0.786&0.849&0.869&0.878&0.862&0.828&0.818\\
        \hline
        Test & Acc& 0.697& 0.757&0.778&0.785&0.776&0.748&0.725\\
        & AUC &0.778&0.842&0.863&0.866&0.853&0.822&0.808\\
        \hline
        \multicolumn{2}{|c|}{{\em CNN-RNN} model} & LST & MST-F & MST-N & MST-SC & SST-1M & SST-C & SST-A \\
        \hline
        Validation & Acc & 0.740& 0.802& 0.816& 0.820& 0.817& 0.801& 0.771\\
        & AUC & 0.819& 0.896& 0.912& 0.912& 0.900& 0.902& 0.861\\
        \hline
        Test & Acc&0.732&0.800&0.816&0.812&0.809&0.796&0.771 \\
        & AUC & 0.815&0.890&0.909&0.902&0.893&0.898&0.862 \\
        \hline
     \end{tabular}}
    \caption{Accuracy and AUC values for the {\em single-tel} and the {\em CNN-RNN} models, for both validation and test datasets, broken down by telescope type.}
    \label{tab:results_summary}
\end{table}

No quality cuts or data preselection were enforced during training, so the models were fed with all images that triggered the telescopes, as opposed to the conventional analysis, where data preselection and quality cuts are routinely performed. In order to illustrate how data preselection cuts affect the learning performance we trained the {\em CNN-RNN} model imposing a telescope multiplicity cut for those events entering the training, namely, we only passed events that triggered a minimum number of telescopes. Results on the validation set demonstrate a substantial improvement in terms of AUC, boosting this metric beyond 0.90 in all telescope designs after a multiplicity cut of 4 triggered telescopes per event is applied (a standard multiplicity cut in the analysis of simulated CTA data~\cite{Acharyya:2019nwy}), with AUC values up to 0.98 for all MST designs and the SST-1M design.

\section{Conclusion and Outlook}
\label{sec:conclusion}

We have presented the CTLearn package, describing its design and configurability. We have shown a usage example in terms of gamma-ray/cosmic-ray IACT event classification performed by two of its built-in DNC-based models. The {\em single-tel} and {\em CNN-RNN} models were trained on datasets of simulated CTA diffuse events. Without performing any quality cuts or data selection, we concluded with AUC values on the test set ranging from 0.78 to 0.87 for the {\em single-tel} model and from 0.82 to 0.91 for the {\em CNN-RNN} model, varying with the telescope design.

The flexibility and configurability of DCN-based algorithms and the rapid evolution of the field open a seemingly infinite number of possibilities for improvement and further development. Some of these areas where development is planned or already ongoing are: building models for full event reconstruction, including the estimation of the energy and the arrival direction of the gamma-ray events, possibly implemented through multitask learning; implementing models that could combine event-level data from a heterogeneous collection of telescope types, enabling IACT-specific metrics and loss functions; and crafting efficient tools to explore the space of hyperparameters for DCN-based models. A comparison between IACT performances obtained with conventional particle classification methods and DCN-based methods is currently in the works. 

\begin{wrapfigure}{R}{0.6\textwidth}
    \vspace{-40pt}
    \includegraphics[width=0.6\textwidth]{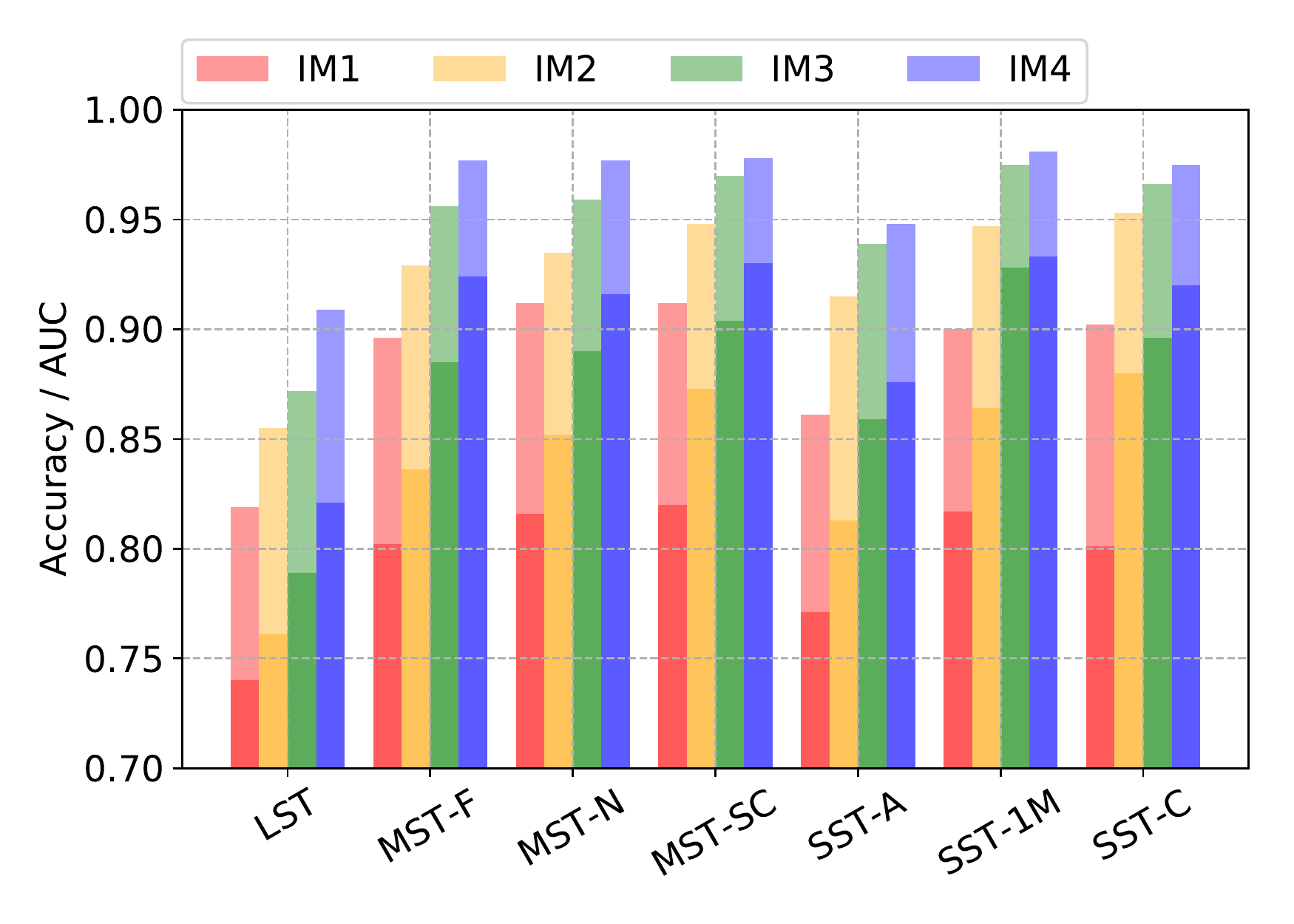} 
    \caption{Validation accuracy (bright colors) and AUC (pale colors) for the {\em CNN-RNN} model, broken down by telescope type and minimum multiplicity cut.}
\end{wrapfigure}

\section{Acknowledgments}
This work was conducted in the context of the CTA Analysis and Simulations Working Group. We thank Johan Bregeon for his support reducing our dataset on the EGI. DN and TM acknowledge support from the former {\em Spanish Ministry of Economy, Industry, and Competitiveness / European Regional Development Fund} grant FPA2015-73913-JIN. AB acknowledges support from NSF award PHY-1229205. BK acknowledges support from NSF awards 1229792 and 1607491. We acknowledge the support of NVIDIA Corporation with the donation of a Titan X Pascal GPU used for part of this research.
\\
\\
This paper has gone through internal review by the CTA Consortium.

\end{document}